\documentclass[aps,preprint]{revtex4}
\usepackage{latexsym}
\usepackage{bbm}
\usepackage{graphics}
\usepackage{amsmath}
\usepackage{amssymb}
\usepackage{graphicx}
\usepackage{graphics}

\newcommand{\beq}{\begin{equation}}
\newcommand{\eeq}{\end{equation}}
\newcommand{\beqa}{\begin{eqnarray}}
\newcommand{\eeqa}{\end{eqnarray}}

\begin{document}

\title{Entanglement enhancement for two spins assisted by two phase kicks}
\author{Gennadiy Burlak$^{1}$, Isabel Sainz$^{2}$, and Andrei B. Klimov$^{2}$$\footnote{klimov@cencar.udg.mx}$}
\affiliation{$^{1}$Centro de Investigaci\'{o}n en Ingenier\'{\i}a y Ciencias Aplicadas,
Universidad Aut\'{o}noma del Estado de Morelos, Cuernavaca, Morelos, M\'{e}xico\\
$^{2}$Departamento de F\'isica, Universidad de Guadalajara, Revoluci\'on
1500, Guadalajara, Jalisco, 44420, M\'exico. }

\begin{abstract}
We study the entanglement dynamics in a two-spin system governed by a
bilinear Hamiltonian and assisted by phase kicks. It is found that the
application of instant kicks to both spins at some specific moments leads to
enhancement of entanglement. This procedure also improves the transient
character of entanglement leading, for large spins, to a formation of a
plateau for the $I$-concurrence. We have numerically investigated the
spin-spin dynamics for several values of spins and observed a substantial
enhancement of entanglement in comparison to the evolution without kicks.
\end{abstract}

\maketitle

\section{Introduction}

Recently, a significant amount of theoretical and experimental research has
been carried out for generating, manipulating and preserving entanglement,
which is the key ingredient for developing quantum technologies. The main
attention is usually paid to studying entanglement properties of qubits
(two-level systems) (see e.g. \cite{qubits,ReviewEnt}), considered as a
basic physical resource. Although, higher dimensional systems (qudits)
could be more appropriate for certain physical applications, entanglement
dynamics, generation and manipulation in higher dimensions is
not as well studied as in two-dimensions.

In particular, it was suggested in \cite{qudits} that it is possible to
increase the security, the bit transmission rate, or both in quantum key
distribution protocols by increasing the dimensionality of the involved
systems. The possibility to perform quantum
computation based on qudits cluster states \cite{cluster1}(has also been noted) . Nevertheless,
just few years ago some studies on entanglement dynamics in higher
dimensions have been done. For example, the dynamics of a initially
entangled two-qudit state was studied in \cite{carvalho}, while in \cite%
{jerzy} it is shown that two non-interacting, initially separable spins can
get entangled via a common purely dephasing environment. More general
results about environment mediated generation of entanglement in higher
dimensions can be found in \cite{benatti}.
%The entanglement power of unitary operations in higher dimensions was also studied, \cite{entpower}.
Other aspects of entanglement in higher dimensional systems, like
entanglement concentration for two qudits in \cite{lucho}, and its
experimental implementation for two qutrits \cite{concen3} were reported. It was also shown in \cite{transfer} that the
quality of entanglement transfer in spin chains actually increases with
the dimension of the spin.

On the other hand, higher dimensional systems (real, as high excited nuclear
spins or effective, as Dicke-like states) are of interest by themselves,
especially in the context of studying the quantum-classical transition, when
the system's dimension becomes sufficiently large. However, it is well-know
that there are no reasons to expect that the classical behavior can be
approached in the large dimension limit for every state \cite{mermin}. In
particular, the coherent spin states (CSS), $\left\vert \zeta
,S\right\rangle $, \cite{coherent} are the most classical ones, and some
specific linear combination of them may reveal non-classical features, like
entanglement, even in the limit of large values of $S$ (see \cite{gerry}).

One of the regular ways for generating and measuring entanglement in many
spin 1/2 system is through spin squeezing (see \cite{SS1} and references
therein), where quantum correlations naturally appear in the basic concept of
squeezed spin states (SSS) \cite{SS}. The simplest Hamiltonian which
produces squeezing starting with coherent states located on the equator of
the Bloch sphere is $H=gS_{z}^{2},$ where $S_{z}=\sum_{j}\sigma _{z}^{(j)}/2$
is a collective spin operator \cite{SS}.

Unfortunately, there are two problems with such dynamical generation of
squeezing and entanglement: a) the squeezing (and entanglement) is always
transient, i.e. it appears periodically
%and its average value is not very high
; b) it does not reach its maximal value, except the simplest two-qubit
case.

In the case of large two spins the situation is rather similar. It is
possible to generate in an easy way the entanglement between these spins,
but it would resemble the above mentioned disadvantages as in the symmetric
combination of many qubits.

In the present article we address the following question: is it possible to
improve the entanglement properties of a two qudit system in a relatively
simple way?

We will show that by applying instant kicks to both spins at some particular
moments we can not only enhance the entanglement, but, more
importantly, essentially improve its transient nature, i.e. reduce the
distance between the maximum and the minimum values, so that the system
evolves into an approximate \textquotedblleft steady\textquotedblright\
state of the interaction Hamiltonian with a high value of entanglement
for sufficiently large spins. We also discuss its possible application to
the purely dephasing environment \cite{jerzy,benatti}.

\section{The model}

Let us consider two spins $S_{1}$ and $S_{2}$, of dimensions $2S_{1}+1$ and $2S_{2}+1$ respectively, which interact according to the Ising-like
Hamiltonian,
\begin{equation}
H=g\hat{S}_{z1}\hat{S}_{z2}.  \label{Hint}
\end{equation}
This kind of interaction was proposed to implement one way quantum
computation with many-level cluster states (see \cite{cluster1} and
references therein). Suppose that each spin is initially in the eigenstate
of  $\hat{S}_{xi}$ with zero phase, $\hat{S}_{xi}\left\vert \zeta _{i}=1,S_{i}\right\rangle
=S_{i}\left\vert \zeta _{i}=1,S_{i}\right\rangle $, which is a CSS placed on
the equator of the Bloch sphere,
\begin{equation}
\left\vert \zeta _{i}=1,S_{i}\right\rangle =\frac{1}{2^{S_{i}}}
\sum_{k_{i}=-S_{i}}^{S_{i}}\sqrt{\frac{(2S_{i})!}{
(S_{i}+k_{i})!(S_{i}-k_{i})!}}\left\vert k_{i},S_{i}\right\rangle ,  \notag
\end{equation}
where $i=1,2$ and $\hat{S}_{zi}\left\vert k_{i},S_{i}\right\rangle
=k_{i}\left\vert k_{i},S_{i}\right\rangle $, so that the joined state for
the system composed by spins $S_{1}$ and $S_{2}$ is $\left\vert \Psi_0\right\rangle =\left\vert \zeta _{1}=1,S_{1}\right\rangle \left\vert
\zeta _{2}=1,S_{2}\right\rangle $ .

Since $\left\vert \Psi_0\right\rangle $ is not an eigenstate of the
Hamiltonian, in the course of evolution the joined state
\begin{equation}
\left\vert \Psi (t)\right\rangle =\hat{U}(t)\left\vert \Psi_0\right\rangle
,\quad \hat{U}(t)=e^{-itg\hat{S}_{z1}\hat{S}_{z2}},  \label{stateNK}
\end{equation}
becomes entangled at some instants. As an entanglement measure we will
use the (normalized) $I$-concurrence \cite{Iconcurrence}, that for pure
states takes the form,
\begin{equation}
\mathcal{C}_{I}=\frac{2S+1}{2S}\left( 1-\text{Tr}\hat{\rho}_{1(2)}^{2}\right) ,\notag
\end{equation}
where $\hat{\rho}_{1(2)}=\text{Tr}_{2(1)}(\hat{\rho}(t))$ is the density
matrix of the composed system $\hat{\rho}(t)=\left\vert \Psi
(t)\right\rangle \left\langle \Psi (t)\right\vert $ traced by the spin $%
S_{2} $ ($S_{1}$), and $S=\min (S_{1},S_{2})$, such that $0\leq \mathcal{C}%
_{I}\leq 1$, being zero for separable states and one for maximally entangled
states.

For the state (\ref{stateNK}), $\mathcal{C}_I=(2S+1)(1-\mathcal{P})/2S$,
where the purity $\mathcal{P}=\text{Tr}\hat{\rho}_{1}^{2}(t)$ is
\begin{equation}
\mathcal{P}=\frac{1}{16^{S_{1}}}\sum_{k,l=-S_{1}}^{S_{1}}\frac{\left[
(2S_{1})!\right] ^{2}\cos ^{4S_{2}}(gt(k-l)/2)}{%
(S_{1}+k)!(S_{1}-k)!(S_{1}+l)!(S_{1}-l)!}.  \notag
\end{equation}

\begin{figure}[tbp]
\includegraphics[width=0.45\textwidth]{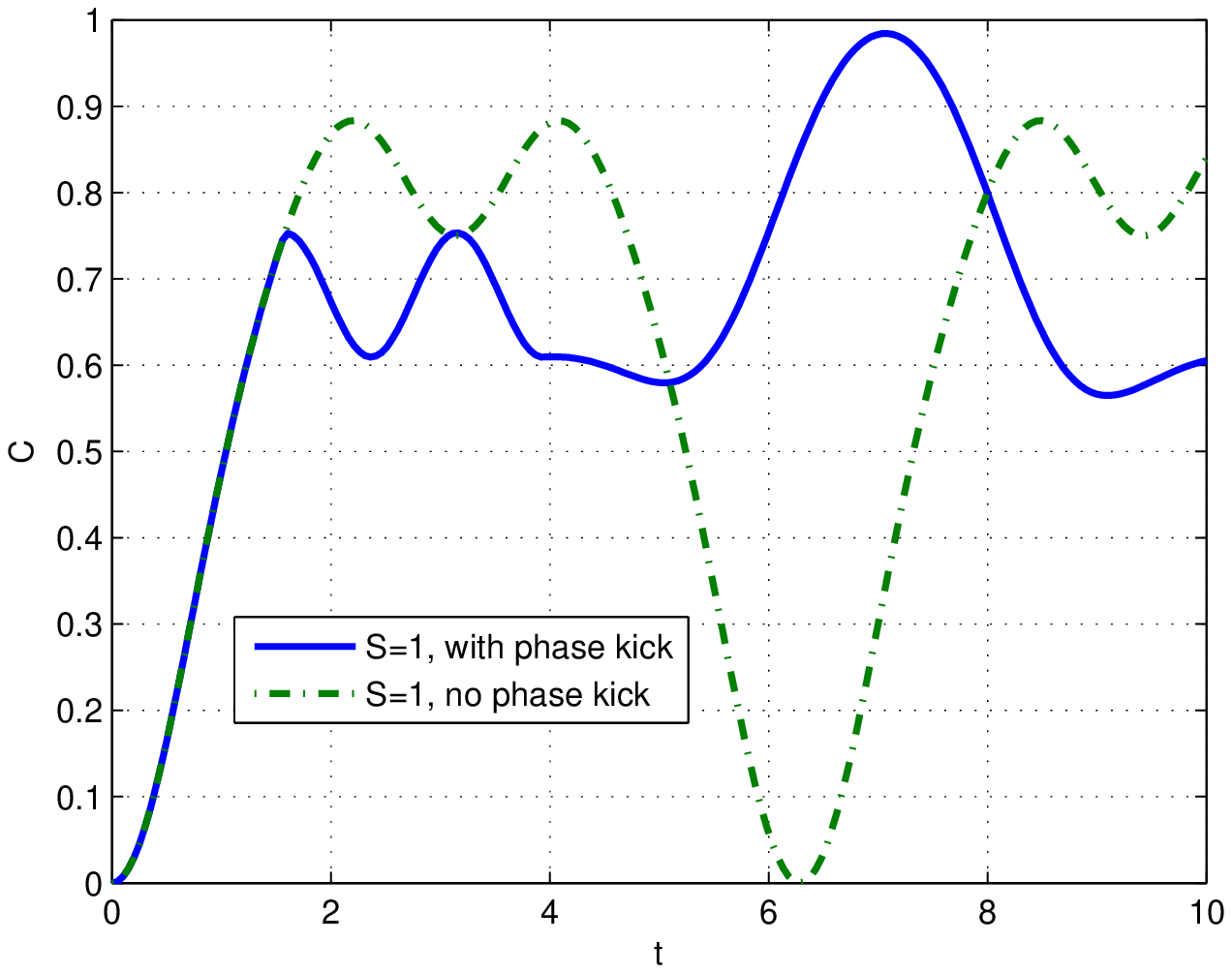}(a)
\includegraphics[width=0.45\textwidth]{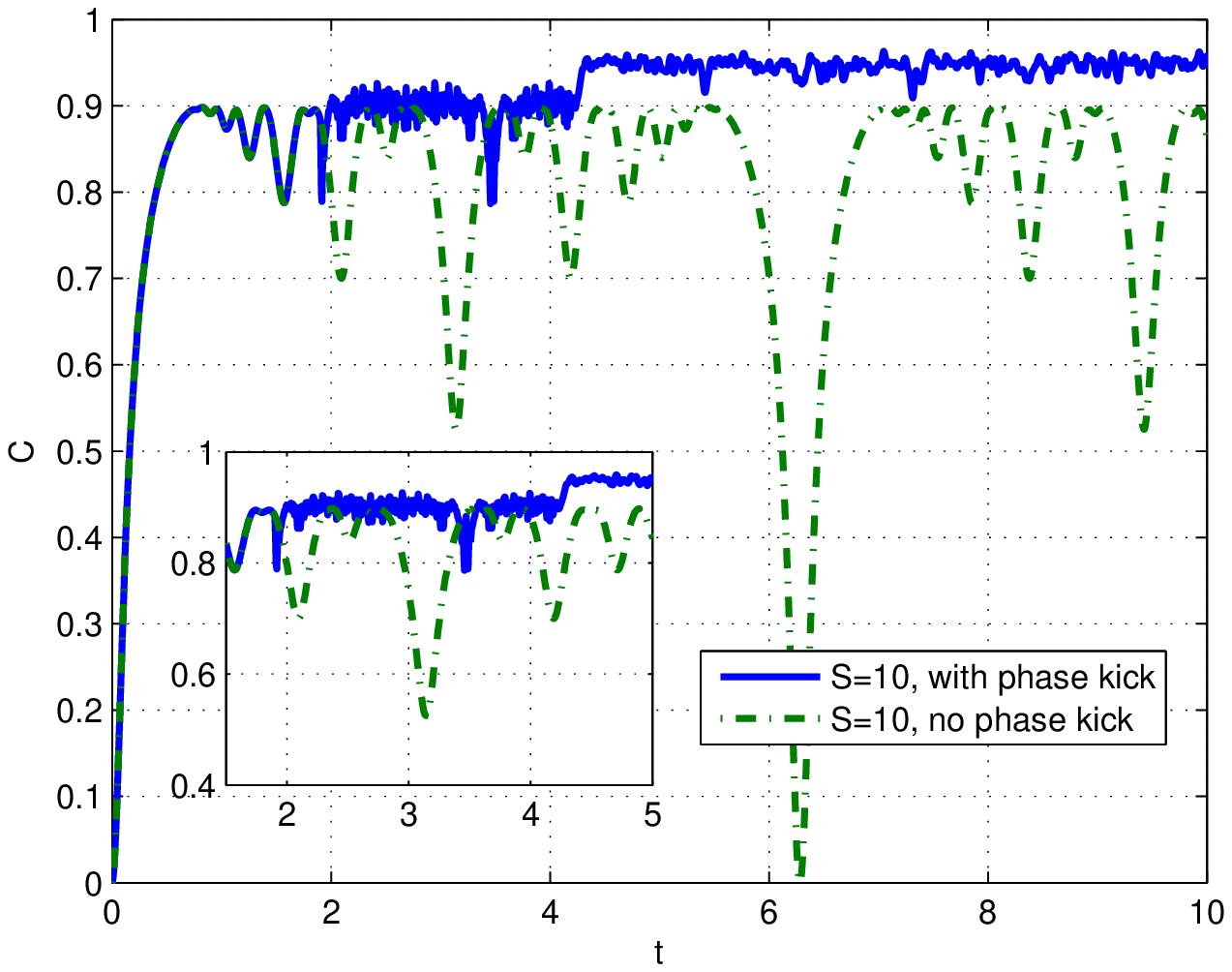}(b)
\caption{(Color online) Time evolution of $\mathcal{C}_I$, without kicks
(green, dash-dotted line) and with phase-kicks (blue, continuous line) for
(a) $S=1$, with the numerical optimization times $t_1=1.6$ and $t_2=3.9$,
and (b) $S=10$, in this case $t_1=1.9$ and $t_2=4.2$ where the inset shows
the details for the second phase kick.}
\label{spin1y10}
\end{figure}

From now on we will focus only on the symmetric case: $S_{1}=S_{2}=S$. Let
us consider the two simplest cases: $S=1/2,1$ . For two qubits, $S=1/2 $, $\mathcal{C}_{I}=\sin ^{2}gt/2$, i.e. it oscillates between zero
and one, so that the spins are maximally entangled at times $gt=(2n+1)\pi $
and they are separable at times $gt=2n\pi $, $n=0,1,\ldots $. Nevertheless,
for $S=1$, the situation is already different, since $\mathcal{C}_I=3\left( 4-2\cos gt-\cos ^{2}gt-\cos ^{4}gt\right)/8$, reaches its maximum
value $\mathcal{C}^{\max}_I\approx 0.88$ at times $gt\approx 2.2,4.1+2n\pi,$
 and thus the spins will never become maximally entangled, although they are
disentangled at $gt=2n\pi,$ $n=0,1,\ldots$. The entanglement dynamics have
a quasi-periodical behavior, as is shown by the (green) dash-dotted line in
Fig. \ref{spin1y10} (a).

In the large spin limit, $S\gg1$, we can approximate the time average of $\mathcal{C}_I$ as follows,
\begin{equation}
\langle \mathcal{C}_{I}\rangle \approx \frac{2S+1}{2S}\left( \frac{2}{\sqrt{2\pi S}}-\frac{1}{2\pi S}\right) ,  \label{Cprom}
\end{equation}
giving $\langle \mathcal{C}_{I}\rangle \approx 0.54$ for $S=1$, which is
quite close to the exact numerical value $\langle \mathcal{C}_{I}\rangle
=0.58$, even if the approximation is done for $S\gg 1$.

\section{Enhancing entanglement}

In order to enhance the maximum achievable entanglement let us apply instant
pulses (kicks) to each spin at certain times $t_{k}$. Such kicks correspond
to rotations around the $y$-axis, and for the $j$-th spin are represented by
the operators $\hat{R}_{j}=e^{-i\pi \hat{S}_{yj}/2}$. The main idea is as
follows: initially both spins are in eigenstates of $\hat{S}_{xi}$,
which can be represented by localized distributions on the equator of the
corresponding Bloch spheres. In the course of evolution governed by Eq.(\ref{Hint}) the state become partially entangled. At some appropriate moment $t_1$ we apply a $\pi /2$-rotation to both spins so that each component of
the angular momentum basis $\left\vert k_i,S\right\rangle $ instantly
transforms into an eigenstate of $\hat{S}_{xi}$. Afterwards, the evolution under
the Hamiltonian (\ref{Hint}) is continued for a while, which continues
coupling both spins. Then, at some time $t_2$ we apply the inverse $\pi /2$
-rotation that transforms again each element $\left\vert k_i,S\right\rangle $
into an eigenstate of $\hat{S}_{xi}$, after that we allow the system to evolve
further:
\beq
\left\vert \Psi (t)\right\rangle =\hat{U}(t-t_{2})\hat{R}^{-1}\hat{U}(t_{2}-t_{1}) \hat{R}\hat{U} (t_{1}-t_{0})\left\vert \Psi_0\right\rangle ,\notag
\eeq
where $\hat{R}=\hat{R}_1\otimes\hat{R}_2$. The instants $t_{1}$ and $t_{2}$ are chosen in order to optimize the
maximum value of achievable entanglement after the second kick. A similar
procedure was proposed to optimally create squeezing in Dicke states \cite{daisuke} and entanglement for continuous-variable systems in \cite{cv1,cv2}.

The reason for applying $\pi /2$-rotations can be explained as follows: we want
to correlate as many components of the spin states as possible, which cannot be done only with the Hamiltonian evolution, since the number of
components of the angular momentum basis, initially involved in the CSS, is
of order $\sqrt{S}$. By applying the kicks at some specific moments we are
able to spread the distribution making it more uniform \cite{cv1, cv2},
while the Hamiltonian evolution produces correlation between spin components
in the angular momentum basis, leading to entanglement.

\begin{figure}[tbp]
\includegraphics[width=.45\textwidth]{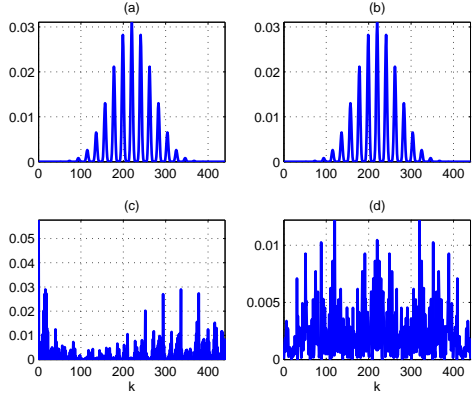}
\caption{(Color online). Distribution of the two-spin probability $\langle\Psi(t)\vert\Psi(t)\rangle$ as function of the number of state $k$ in the long
basis $\{0,\ldots,(2S+1)(2S+1)\}$ for $S=10$. (a) for the initial state, $t=0 $, (b) before the first phase kick at $t<t_{1}$, (c) after the first kick, but
before the second kick at $t_{1}<t<t_{2}$, and (d) after the second kick $t>t_{2}$.}
\label{distEdos}
\end{figure}

In Fig. \ref{distEdos}, we plot the distribution of the two-spin probability
$\langle \Psi (t)|\Psi (t)\rangle $ as a function of the state number $k$,
corresponding to the basis $\left\vert k_{1},10\right\rangle \left\vert
k_{2},10\right\rangle $, in the following way, $\left\vert
-10,10\right\rangle \left\vert -10,10\right\rangle $ corresponds to $k=0$,
the next state $\left\vert -10,10\right\rangle \left\vert -9,10\right\rangle
$ corresponds to $k=1$, and so on. For the initial state (Fig. \ref{distEdos}
(a)), the distribution is centered on $\left\vert 0,10\right\rangle
\left\vert 0,10\right\rangle $, that is, $k=2S(S+1)+1=221$. The Hamiltonian
evolution does not affect the distribution of the amplitudes of the
components of the basis $k$ ($\left\vert k_{1},S\right\rangle \left\vert
k_{2},S\right\rangle $) as shown by Fig. \ref{distEdos} (b). It can be
seen in Fig. \ref{distEdos} (c) that after the first instantaneous
rotation of each spin, the distribution of $\langle \Psi (t)|\Psi (t)\rangle
$ spreads over the states $k$, this effect is increased after the second
instantaneous rotation of each spin (Fig. \ref{distEdos} (d)).

For semi-integer spins ($S=(2n+1)/2$, $n=0,1,\ldots $), the situation is
slightly different: the initial distribution is similar to Fig. \ref{distEdos} (a), but instead of having a single maximum at the center, it has
two maxima, which leads to non-essential (but observable) differences in the
dynamics.

The analytical expressions are very cumbersome for higher spins, so that we
have to use numerical optimization for $S\geq 3/2$.

As noted before, for the simplest case, when $S=1/2$, the system
oscillates between the separable and maximally entangled states and the
kicks produce just a phase shift, that does not affect the dynamics.

For $S=1$, we have been able to analytically optimize $\mathcal{C}_I$, and
found that the entanglement acquires its maximum value when the kicks are
applied at times $t_{1}\approx 1.6$ and $t_{2}\approx 3.9$ (which is also in
accordance with the numerical results). For these times, $\mathcal{C}_I$
reaches the maximum $\mathcal{C}_{I}^{\max }\approx 0.98$ at times $t\approx
7.1+2\pi n$, which should be compared with that obtained by the pure
Hamiltonian evolution, where $\mathcal{C} _{I}^{\max }\approx 0.88$.
However, it is more important that after the second kick the
quasi-periodical behavior of $\mathcal{C}_I$ has the minimal value $\mathcal{C}_{I}^{\min }\approx 0.57$, which means that the spins are
entangled at any time $t>t_{2}>t_{1}$, as shown in Fig. \ref{spin1y10}
(a), and $\langle\mathcal{C}_I\rangle$ increases from $\langle \mathcal{C}_{I}\rangle \approx 0.58$ to $\langle \mathcal{C}_{I}\rangle \approx 0.71$.

\begin{figure}[tbp]
\includegraphics[width=.45\textwidth]{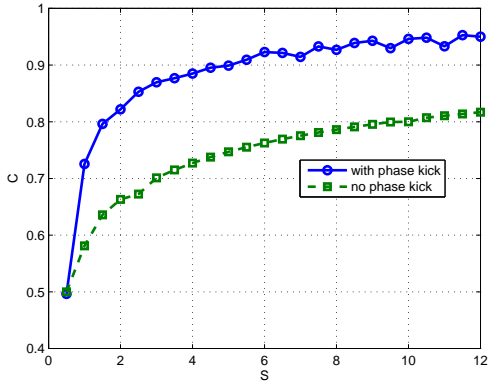}
\caption{(Color online) The behavior of $\langle\mathcal{C}_I\rangle$
against the spin number $S\leq12$. The (green) squares joined by the dashed
line correspond to the two-spin system under the Hamiltonian evolution given
in Eq. \protect\ref{Hint}. The (blue) circles joined by a continuous line
correspond to the enhanced behavior, for times $t\geq t_2$.}
\label{concurrencias}
\end{figure}

In Fig. \ref{spin1y10} (b), we plot the entanglement dynamics for $S=10$. In
this case, the effect of enhancing the maximum of entanglement for $t>t_{2}$, is not that pronounced as for $S=1$, it increases from $\mathcal{C}%
_{I}^{\max }\approx 0.90$ to $\mathcal{C}_{I}^{\max }\approx 0.96$. However,
another important effect emerges: the transient nature of entanglement becomes
significantly improved, i.e. after the second kick $\mathcal{C}_I$ rapidly
oscillates with a very small amplitude around its average value $\langle
\mathcal{C}_{I}\rangle \approx 0.95$. So that the spins are entangled for
all times $t>t_{2}$, with a high value. On the other hand, we observe, that
in the case of pure Hamiltonian evolution $\mathcal{C}_I$ periodically
oscillates between its maximum $\mathcal{C}_{I}^{\max }\approx 0.90$ and
zero.

The average $\mathcal{C}_I$ grows slowly as $S$ increases, and even if
the maximum and minimum are getting closer, the rate of increase of the average is not as fast as could be expected. This is
a consequence of the growing character of $\langle \mathcal{C_{I}}\rangle $
for the Hamiltonian dynamics for large $S$ as seen from Eq. (\ref{Cprom}). To represent such behavior clearly we plot in Fig. (\ref{concurrencias}) $\langle\mathcal{C}_I\rangle$ for the Hamiltonian evolution (squares) and
after applying the sequence of kicks (circles), numerically obtained.

\section{Discussion}

We have studied the dynamics of a two-spin entanglement assisted by
two phase kicks. The instantaneous kicks were applied at times such
that the maximum of $\mathcal{C}_I$ has been (numerically) optimized. We
found that such kicks not only lead to an enhancement of the two-spins
maximum entanglement, but also to a substantial increasing of the minimum
and average values of $\mathcal{C}_I$ for times after the second kick. Such
an effect becomes much more pronounced for large spins $S\gg 1$, when $\mathcal{C}_I$ tends to a constant value very close to unity.

It is worth noting that the interaction Hamiltonian similar to (\ref{Hint})
appears in several effective processes. In particular, when two
non-interacting spins are embedded in a dephasing environment \cite{jerzy}.
The Hamiltonian describing such evolution has the form $H=\Omega
_{1}S_{z1}+\Omega _{2}S_{z2}+\sum_{k}\omega _{k}a_{k}^{\dag
}a_{k}+(S_{z1}+S_{z2})\sum_{k}g_{k}(a_{k}+a_{k}^{\dag })$, such that $\omega
_{k}\gg \Omega _{1,2}$, $g_{k}$. The fast field is effectively decoupled
from the spin subsystem and thus can be adiabatically eliminated in a
standard way \cite{SR} by applying a small unitary transformation $U=\exp {[(S_{z1}+S_{z2})\sum_{k}\epsilon _{k}(a_{k}^{\dag }-a_{k})]}$, ${\epsilon }_{k}=g_k/\omega _{k},$ so that the effective Hamiltonian, $H_{eff}=UHU^{\dagger }$ (diagonal on the field variables) takes the form,
\begin{eqnarray*}
H_{eff} &=&(\Omega _{1}-\epsilon S_{z1})S_{z1}+(\Omega _{2}-\epsilon
S_{z2})S_{z2} \\
&&+\sum_{k}\omega _{k}a_{k}^{\dag }a_{k}-2\epsilon S_{z1}S_{z2},
\end{eqnarray*}
where $\epsilon =\sum_{k}g_k\epsilon _{k}$. It can be observed that the effect
of bosonic bath in this case is reduced to an effective interaction between
spins, which may give rise to a creation of spin entanglement and a quadratic phase shift
that is not important for entanglement generation.

\section{Acknowledgements}

The work of G.B is partially supported by CONACyT grant 47220. The work of
I.S. was supported by CONACyT grant 74897.

\end{document}